# Ultrasensitive Room-Temperature $NO_2$ Gas Sensor Based on $In_2O_3$/$NbS_2$ Heterojunction

P K Shihabudeen [a$], Alex Sam [b$], Shih-Wen Chiu [c], Ta-Jen Yen [b], Kea-Tiong Tang [a,*]

[a] Department of Electrical Engineering, National Tsing Hua University, Hsinchu, Taiwan,

[b] Department of Materials Science and Engineering, National Tsing Hua University, Hsinchu, Taiwan,

[c] Enosim Bio-tech Co., Ltd., Hsinchu, Taiwan

**Abstract**

Niobium disulfide ($NbS_2$), a two-dimensional transition metal dichalcogenide with semi metallic conductivity and high surface activity, offers promising properties for electronic and sensing applications. In this study, we report a high-performance $NO_2$ gas sensor based on a heterostructure comprising a spin-coated $In_2O_3$ film on a semi-metallic $NbS_2$ film. The $NbS_2$ was synthesized via chemical vapor deposition (CVD) and subsequently transferred to a substrate prior to $In_2O_3$ coating. Pristine $In_2O_3$ exhibited limited gas response, and $NbS_2$ alone was inert to $NO_2$; however, the $NbS_2$/$In_2O_3$ heterostructure demonstrated a significant enhancement in sensing performance. This included a high response of 7520 % at 500 ppb and 46.5 % at 10 ppb, along with reasonable response (~22 s) and recovery (~155 s) times. The sensor maintained robust performance across a wide humidity range (25–90% RH), showed excellent selectivity against interfering gases, and remained stable over one month of operation. These findings highlight the potential of

* Corresponding author: Electronic mail: kttang@ee.nthu.edu.tw
$ Both authors contributed equally to this work.

semimetal/semiconductor heterojunctions for developing room-temperature gas sensors with high sensitivity and environmental stability

**Introduction**

Two-dimensional (2D) transition metal dichalcogenides (TMDCs) have gained significant attention in recent years due to their exceptional electrical, optical, and mechanical properties [1–3]. While semiconducting TMDCs such as $MoS_2$ (Molybdenum disulfide) and $WS_2$ (Tungsten disulfide) have been extensively explored for applications in transistors, photodetectors, and gas sensors, metallic TMDCs are now emerging as promising materials for contact engineering and electronic interfaces [4–7]. In particular, semi-metallic niobium disulfide ($NbS_2$) has attracted growing interest owing to its high electrical conductivity, van der Waals surface, and compatibility with other 2D semiconductors [8–10]. These attributes make $NbS_2$ an excellent candidate for forming low-resistance heterojunctions with semiconducting oxides, potentially enabling improved charge transfer and sensing performance [11–13].

$NbS_2$ exists in three polymorphs, 3R (rhombohedral), 1T (Trigonal) and 2H (hexagonal), exhibiting metallic and superconducting behavior [7,14,15]. While mechanical exfoliation provides access to few-layer $NbS_2$, it offers limited control over flake size and thickness, making it less suitable for scalable device fabrication. Recent efforts using chemical vapor deposition (CVD) have achieved the growth of all phases $NbS_2$ [8,9]. In addition to its metallic nature, $NbS_2$ exhibits intriguing physical phenomena such as charge-density wave (CDW) behavior and demonstrates high electrocatalytic activity toward the hydrogen evolution reaction (HER) [10,16]. These unique properties make $NbS_2$ not only a valuable platform for exploring fundamental phenomena but also a versatile material for functional heterostructures. $NbS_2$ is often studied alongside semiconducting TMDCs due to its favorable band alignment, which enables the formation of low Schottky barriers at their interfaces [12,17].

Other 2D metal sulfides, like $MoS_2$ and $WS_2$, have been widely studied for gas sensing applications and exhibit gas response at room temperature [18,19]. However, their practical performance is often limited by slow recovery kinetics at room temperature, inferior selectivity, and poor stability. Recent theoretical studies have identified $NbS_2$ as an excellent candidate for $NO_2$ sensing, predicting high sensitivity and rapid recovery characteristics. There are only a few reports available on experimental studies on $NbS_2$-based gas sensors. Kim et al. reported the room-temperature $NO_2$ sensing properties of 2D $NbS_2$ nanosheets and, through first-principles density functional theory calculations, demonstrated that edge configurations, determined by synthesis conditions, significantly influence sensing performance [11]. However, their sensor exhibited low sensitivity and slow response and recovery times. In another study, Azizi et al. explored the potential of $Re_{0.5}Nb_{0.5}S_2$ for $NO_2$ sensing at room temperature [20]. While their sensor showed high selectivity and good stability under varying humidity conditions, the reported limit of detection was above 50 ppb, which is inadequate for air quality monitoring applications that require detection in the 0–100 ppb range.

Semiconducting metal oxides (SMOs) have been widely reported to enhance the gas sensing performance of other metal chalcogenides [18]. Among them, indium oxide ($In_2O_3$) is one of the most commonly used SMOs for $NO_2$ sensing due to its excellent chemical and thermal stability [21,22]. $In_2O_3$ can form heterojunctions with other materials, creating charge accumulation regions at the interface that serve as active sites for gas adsorption, thereby significantly improving sensing performance [18,23,24]. In this work, we exploit the semi-metallic nature of $NbS_2$ in combination with In$_2$O$_3$ to construct a heterostructure-based NO$_2$ gas sensor capable of efficient operation at room temperature. Our results show that electron accumulation at the NbS$_2$/In$_2$O$_3$ interface plays a key role in achieving high sensitivity and stable sensing characteristics under ambient conditions.

**Experimental**

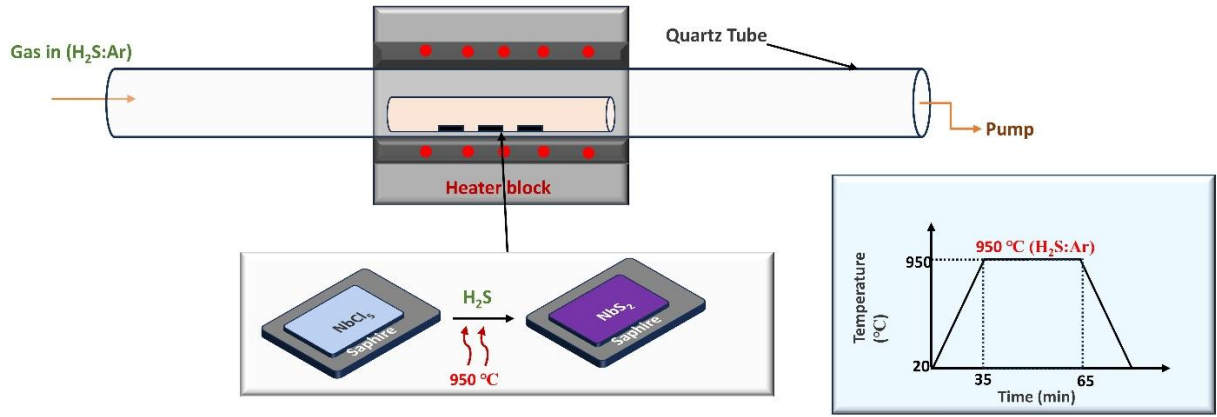

Figure 1. Schematic representation of the CVD process used for the synthesis of NbS$_2$ thin film

The NbS$_2$ thin film was synthesized using a chemical vapor deposition (CVD) process on c-plane sapphire (C-Al$_2$O$_3$) substrates. Prior to growth, the substrates were sequentially cleaned in acetone, isopropanol (IPA), and deionized (DI) water using ultrasonication. Cleaned wafers were then annealed at 300 °C for 3 hours in ambient air to improve surface cleanliness. To initiate synthesis, the sapphire substrates were spin-coated with a 1 mM NaCl aqueous solution, followed by a solution of NbCl$_5$ dissolved in IPA. After drying, the coated substrates were loaded into a quartz tube placed inside a horizontal tube furnace. The CVD reactor was purged with high-purity argon gas for 1 hour to eliminate residual oxygen and moisture. During the growth process, hydrogen disulfide (H$_2$S) gas was introduced at a flow rate of 20 sccm, using argon as a carrier gas. The growth was carried out at 950 °C under a pressure of 200 Torr for 30 minutes. After the reaction, the furnace was allowed to cool naturally to room temperature. To mitigate sulfur vacancy formation in the NbS$_2$ films, the H$_2$S flow was continued at a reduced rate of 10 sccm during the cooling stage until the furnace temperature dropped below 150 °C. The overall synthesis process and reactor setup are illustrated in Figure 1.

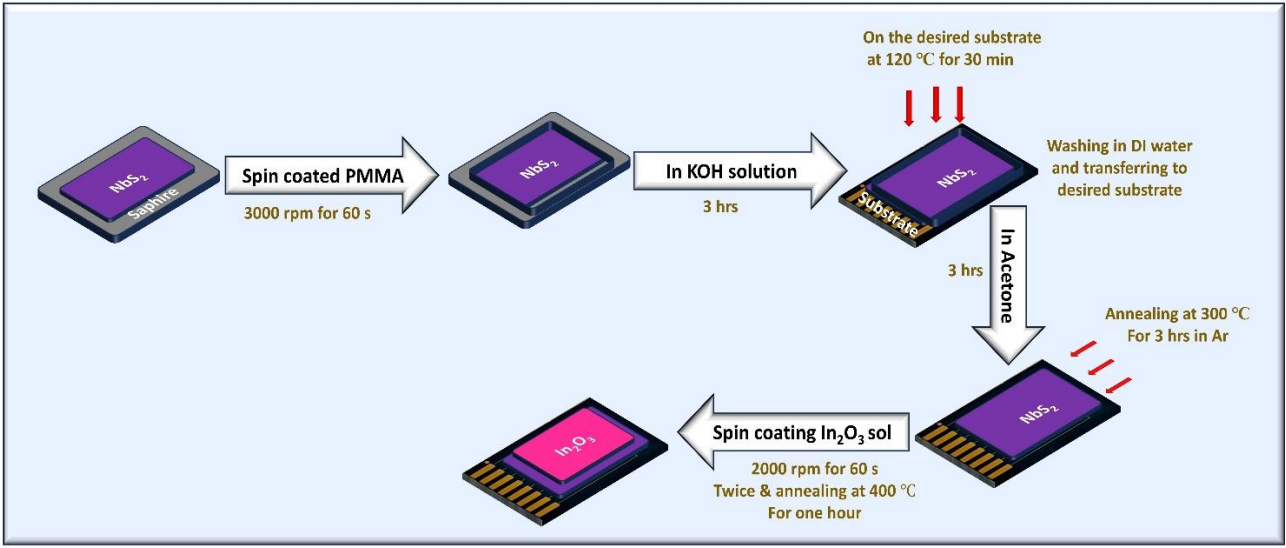

Figure 2. Schematic illustration of the step-by-step fabrication process of the gas sensor.

Following CVD synthesis, the NbS$_2$ thin film was transferred from the c-sapphire growth substrate to the desired target substrate using a PMMA-assisted wet transfer method, as illustrated in Figure 2. A SiO$_2$/Si substrate patterned with seven microsensor arrays (400 × 400 µm²) and interdigitated gold electrodes (electrode spacing: 20 µm; width: 25 µm) was used as the target substrate. Initially, a layer of polymethyl methacrylate (PMMA) was spin-coated onto the NbS$_2$ film at 3000 rpm for 60 seconds and dried under ambient conditions overnight. To initiate delamination, one edge of the PMMA layer was gently scratched to allow infiltration of the etchant. The sample was immersed in a 1 M potassium hydroxide (KOH) solution for 3 hours to release the PMMA/ NbS$_2$ stack. The detached film was rinsed thoroughly with deionized water and transferred onto the target substrate. The sample was baked at 120 °C for 30 minutes to improve adhesion, and the PMMA layer was removed by immersion in acetone for 3 hours. Finally, the transferred film was annealed at 300 °C in an argon atmosphere to enhance film quality and interfacial contact

To fabricate the $NbS_2/In_2O_3$ heterostructure sensor, a sol-gel solution of indium oxide was prepared by dissolving 1.45 g of indium acetate in 20 mL of absolute ethanol (99.9%). Ethanolamine (≥98%) was added dropwise under continuous stirring at 60 °C until a clear and homogeneous solution was obtained. The stirring was continued for 4 hours to ensure solution stability. The resulting sol was spin-coated twice onto the transferred $NbS_2$ film at 2000 rpm for 60 seconds. After each coating cycle, the film was pre-baked at 80 °C on a hot plate to remove residual solvents. The coated films were then annealed in air at 400 °C for 1 hour to form a uniform $In_2O_3$ layer and complete the sensor fabrication. For comparison, control sensors based on bare $In_2O_3$ and bare $NbS_2$ were also fabricated following similar procedures to evaluate the individual gas sensing performance of each material.

**Materials characterization**

The structural properties of the synthesized thin films were examined using X-ray diffraction (XRD) with a Bruker D2 Phaser diffractometer employing Cu Kα radiation (λ = 1.5406 Å). The surface morphology and topography were characterized by field-emission scanning electron microscopy (FE-SEM, JEOL JSM-8010). Raman spectroscopy was carried out using a HORIBA HR800 spectrometer with a 532 nm laser excitation source to probe the vibrational modes and confirm phase purity. To investigate the chemical states and electronic structure of the materials, X-ray photoelectron spectroscopy (XPS) and ultraviolet photoelectron spectroscopy (UPS) measurements were performed using an ULVAC-PHI PHI 5000 Versa probe II system, equipped with a monochromatic Al Kα X-ray source (1486.6 eV) for XPS and a He I (21.2 eV) UV source for UPS. Cypher Asylum Research System was used to perform Scanning Kelvin Probe Microscopy (SKPM) for both AFM and KPFM analysis. A conductive Ti/Ir-coated tip was employed, with the tip bias voltage set to 3 V.

**Sensor measurements**

The sensing setup used in the present study is illustrated in Figure S1. The fabricated sensor was mounted on a heating unit within a sealed sensing chamber, where the operating temperature was maintained at 25 °C (room temperature). Electrical contact between the sensor and the measurement circuit was established using spring-loaded contact pins. An air compressor supplied dry air (25 °C, 25% RH), while $NO_2$ and other test gases were introduced into the chamber from certified gas cylinders. To achieve various $NO_2$ concentrations, a set of mass flow controllers (MFCs) was employed. By inputting the desired gas concentration into the control system, the MFCs automatically adjusted the flow rates of $NO_2$ and air, maintaining a constant total flow rate of 400 standard cubic centimeters per minute (sccm). Relative humidity was controlled using a water column, where air was bubbled through water before mixing. The height of the water column determined the final RH level: higher columns corresponded to increased humidity. After humidification, the air was mixed with the target gas in a gas mixer and then introduced into the sensing chamber. A humidity sensor inside the chamber continuously monitored the RH during experiments.

For electrical measurements, the sensor was connected in series with a load resistor of comparable resistance, and a constant 5 V bias was applied using a DC power supply. The output voltage was recorded using a data acquisition system (DAQ, NI USB-6009) controlled via LabVIEW software.

The sensor response, S was defined as:

$$S = \frac{R_g}{R_a} \tag{1}$$

where $R_a$ is the steady-state resistance in air and $R_g$ is the resistance in the presence of target gas. Response and recovery times were calculated as the durations required for the sensor to reach 90% of the total change in resistance upon exposure to or removal from the test gas, respectively. Sensor recovery was facilitated by exposing the device to ambient air post-exposure to restore its baseline resistance.

**Result and discussion**

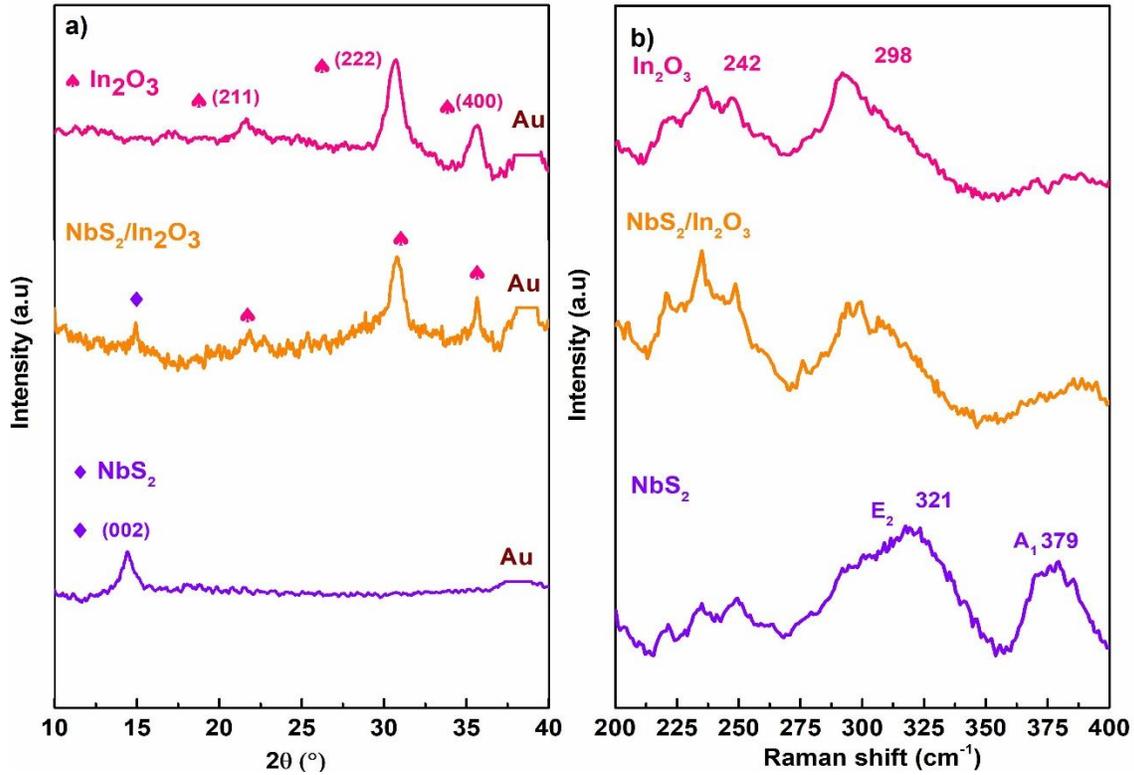

Figure 3. (a) XRD patterns and (b) Raman spectra of the synthesized thin films, illustrating the crystalline structure and vibrational characteristics of each film

To confirm the successful formation and crystalline quality of $NbS_2$, $In_2O_3$, and their heterostructure, X-ray diffraction (XRD) and Raman spectroscopy analyses were performed, as shown in Figure 3. The XRD scans were conducted over a 2θ range of 10° to 40°, as shown in figure 3(a), which effectively covers the prominent diffraction peaks of the materials under investigation. All samples exhibited well-defined crystalline peaks, indicating successful formation of the desired phases. For the $NbS_2$ thin film, a single prominent diffraction peak corresponding to the (002) plane was observed. This peak is characteristic of the 2H-type stacking sequence in layered $NbS_2$, and its presence confirms the formation of a crystalline structure along the c-axis [25]. The lack of additional peaks within the scanned range suggests a preferential orientation or along the [002] direction, which is common in van der Waals layered materials.

The XRD pattern of the $In_2O_3$ film revealed multiple sharp peaks corresponding to the (211), (222), and (400) planes, which are attributed to the cubic phase of indium oxide. This confirms the formation of a polycrystalline film with good crystallinity and phase purity. In the case of the heterostructure, peaks corresponding to both $NbS_2$ and $In_2O_3$ were clearly identified, indicating that the heterostructure preserves the individual crystalline characteristics of both materials. Importantly, no additional or unidentified peaks were present in any of the XRD patterns, suggesting the absence of secondary phases or undesired chemical interactions during film formation.

To further confirm the structural identity of $NbS_2$ and assess the vibrational characteristics of the films, Raman spectroscopy was performed (Figure 3. (b)). Layered transition metal dichalcogenides like $NbS_2$ are known to exhibit four Raman-active modes, with the most prominent being the in-plane E modes and out-of-plane A modes. In the Raman spectrum of $NbS_2$, two distinct peaks were observed at 321 cm$^{-1}$ and 379 cm$^{-1}$, corresponding to the $E_2$ (in-plane) and $A_1$ (out-of-plane) vibrational modes, respectively [14]. These modes are signatures of the 2H-type $NbS_2$ phase, providing additional confirmation of the material's structural configuration and crystallinity.

In the heterostructure, these characteristic NbS$_2$ peaks remained visible, indicating that the integration with In$_2$O$_3$ did not disrupt the layered structure of NbS$_2$. Additionally, a peak at 298 cm$^{-1}$ was observed in both the In$_2$O$_3$ film and the heterostructure, which is attributed to the bending vibration of the InO$_6$ octahedra, a vibrational mode commonly associated with cubic In$_2$O$_3$ [26]. The preservation of individual Raman signatures of both NbS$_2$ and In$_2$O$_3$ in the heterostructure highlights the structural compatibility and successful formation of the layered hybrid architecture.

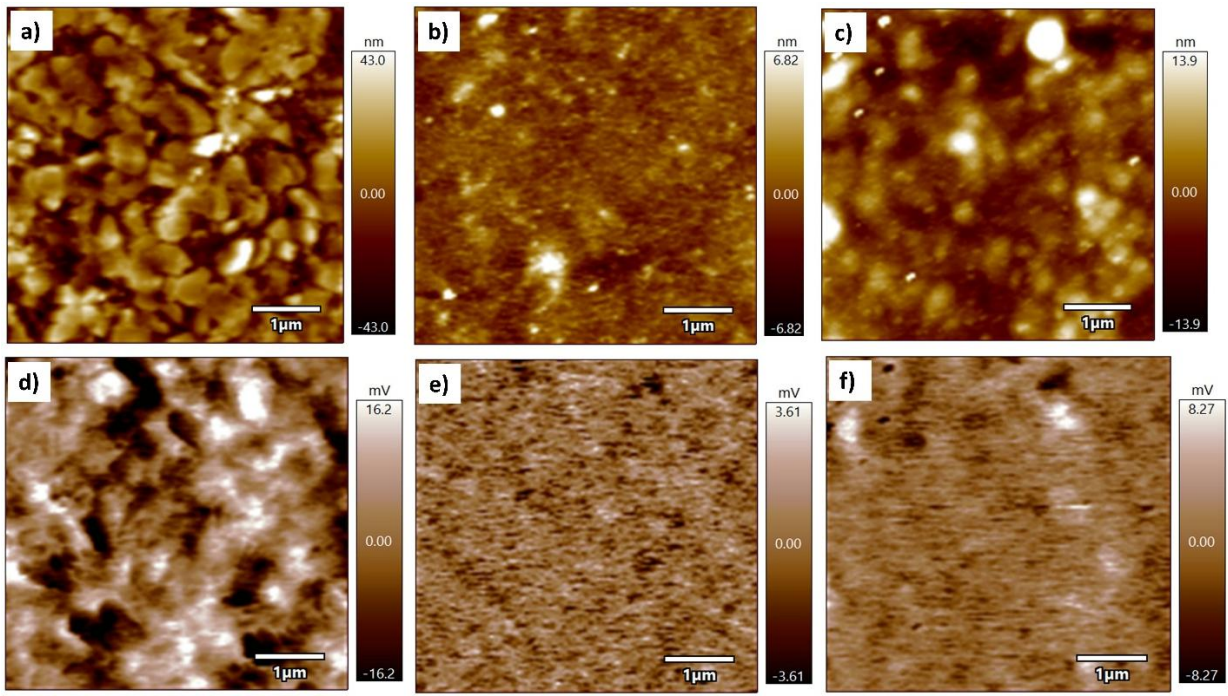

Figure 4. AFM topography (a–c) and surface potential (d–f) images of (a, d) NbS$_2$, (b, e) In$_2$O$_3$, and (c, f) the NbS$_2$/In$_2$O$_3$ heterostructure film

The figure 4 presents the AFM and corresponding KPFM images of NbS$_2$, In$_2$O$_3$, and their heterostructure thin films. Figure 4(a), (b), and (c) show the AFM topography images of NbS$_2$, In$_2$O$_3$, and the NbS$_2$/ In$_2$O$_3$ heterostructure, respectively. The NbS$_2$ film exhibits a relatively rough surface with an average roughness of 13.5 nm, indicating a textured morphology. In contrast, the

$In_2O_3$ film shows a smooth and compact surface with a much lower roughness of 1.45 nm. The heterostructure film displays an intermediate surface morphology, with a reduced roughness of 4.6 nm, suggesting that the $In_2O_3$ layer partially smooths the underlying $NbS_2$ surface. Figure 4(d), (e), and (f) display the corresponding KPFM surface potential images of the same films. The $NbS_2$ film exhibits the highest surface potential due to its semi metallic nature, while $In_2O_3$ shows the lowest, consistent with its semiconducting properties. The heterostructure presents an intermediate surface potential, indicating charge redistribution at the interface. Additionally, $NbS_2$ displays notable electronic inhomogeneity, likely arising from variations in local grain structure or surface states, whereas $In_2O_3$ shows a uniform surface potential distribution, reflecting consistent electronic behavior. The moderate surface potential contrast observed in the heterostructure further confirms interfacial interaction between $NbS_2$ and $In_2O_3$. These findings collectively validate the formation of a heterostructure with improved surface uniformity and modulated electronic properties. The thickness of the $NbS_2$ film was also analyzed using AFM, as shown in Figure S2 in the supplementary data. The step-height measurement indicates that the film has a thickness of approximately 10 nm, confirming the formation of a uniform and ultrathin $NbS_2$ layer suitable for heterostructure fabrication.

The film morphology was further examined using SEM, as presented in Figure S3(a-c) of the supplementary data. Consistent with the AFM results, the $In_2O_3$ and heterostructure films exhibit smoother and more uniform surfaces compared to the relatively rough $NbS_2$ film. Additionally, the TEM image shown in Figure S3(d) reveals a single triangular flake of $NbS_2$, with clearly visible lattice fringes. These fringes provide further evidence of the crystalline nature of the $NbS_2$ film.

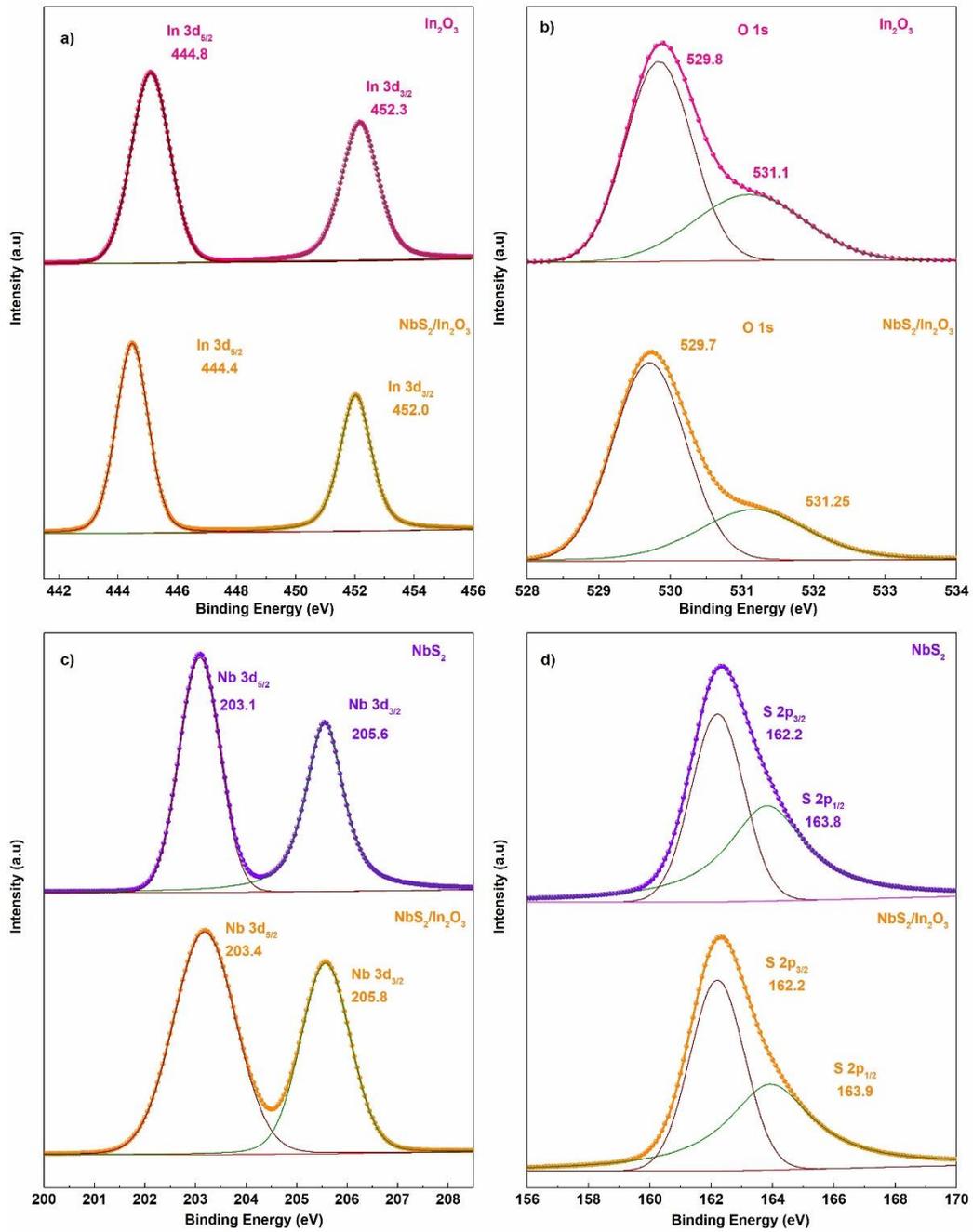

Figure 5. High-resolution XPS spectra of the thin films: (a) In 3d, (b) O 1s, (c) Nb 3d, and (d) S 2p core levels, highlighting the elemental composition and chemical states in the respective samples.

To analyze the chemical nature of the sensor films, X-ray photoelectron spectroscopy (XPS) was performed. Figure 5 presents the high-resolution spectra of In 3d, O 1s, Nb 3d, and S 2p for the three films. Figure 5(a) shows the In 3d doublet peaks for both the pristine $In_2O_3$ and the heterostructure films. These peaks, corresponding to In $3d_{5/2}$ and In $3d_{3/2}$, appear at 448.8 eV and 452.3 eV for $In_2O_3$, and at 444.4 eV and 452 eV for the heterostructure, respectively [18,26]. The observed shift to lower binding energy in the heterostructure suggests electron transfer from $NbS_2$ to $In_2O_3$.Further, Figure 5(b) displays the O 1s spectra for both $In_2O_3$ and the heterostructure. The peaks were deconvoluted and identified at 529.8 eV and 531.0 eV for $In_2O_3$, and at 529.7 eV and 531.1 eV for the heterostructure. The lower binding energy peak corresponds to the In-O lattice bond, while the higher binding energy peak is attributed to non-stoichiometric oxygen species, which may include adsorbed oxygen, oxygen vacancies, or hydroxyl groups [27,28].

Moving on, the Nb 3d spectra are shown in Figure 5(c). The Nb 3d doublet peaks, corresponding to Nb $3d_{5/2}$ and Nb $3d_{3/2}$, are observed at 203.1 eV and 205.6 eV for the pristine $NbS_2$ film, and at 203.4 eV and 205.8 eV for the heterostructure [11,16]. A slight shift to higher binding energy after heterostructure formation further supports the occurrence of electron transfer from $NbS_2$ to $In_2O_3$.Finally, the S 2p spectra for both $NbS_2$ and the heterostructure are illustrated in Figure 5(d). The spectra are deconvoluted into the S $2p_{3/2}$ and S $2p_{1/2}$ peaks. These appear at 162.2 eV and 163.8 eV for $NbS_2$, and at 162.2 eV and 163.9 eV for the heterostructure, indicating minimal chemical shift and confirming the chemical stability of the sulfide component during heterostructure formation [11,16].

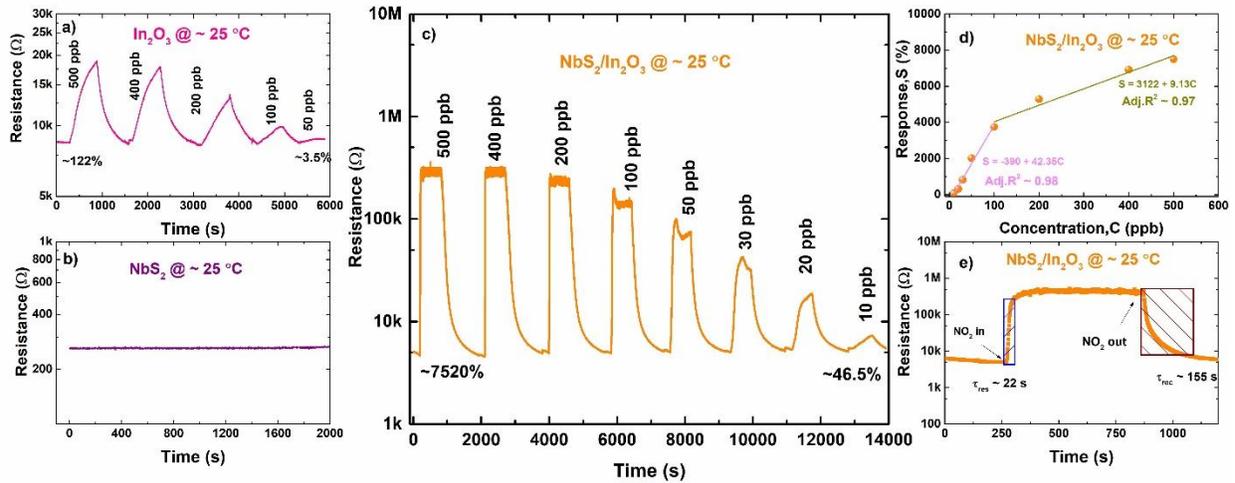

Figure 6. NO$_2$ sensing characteristics of (a) In$_2$O$_3$, (b) NbS$_2$, and heterostructure films; (d) sensitivity curve of the heterostructure film; and (e) resistance transient showing the response and recovery behavior of the heterostructure film

Figure 6 illustrates the NO$_2$ sensing characteristics of three different sensing configurations, pristine In$_2$O$_3$, NbS$_2$, and NbS$_2$/In$_2$O$_3$ heterostructure films, evaluated across a concentration range of 10 to 500 ppb. The gas exposure time for each measurement was fixed at 600 seconds. The pristine In$_2$O$_3$ sensor (Figure 6(a)) demonstrated a measurable response to NO$_2$ at concentrations as low as 50 ppb, with a response of 3.6%. The response increased with concentration, reaching 122% at 500 ppb. Although the response continued to rise without saturation during the exposure period, the sensor showed reasonable recovery toward the baseline after gas removal, suggesting a reversible sensing.

In contrast, the NbS$_2$ film (Figure 6 (b)) exhibited a low baseline resistance (~250 Ω), in agreement with its semi metallic character. However, it failed to show any detectable response to NO$_2$, even at concentrations as high as 15 ppm. This behavior suggests a weak interaction between NO$_2$ molecules and the NbS$_2$ surface, likely due to limited active adsorption sites or minimal charge transfer. This result indicates that pristine NbS$_2$ alone is not suitable for gas sensing under the given conditions. However, when combined with In$_2$O$_3$ to form a heterostructure, the gas sensing properties improved dramatically. As shown in Figure 6(c), the NbS$_2$/In$_2$O$_3$ heterostructure exhibited a highly enhanced and nonlinear sensing response across the tested concentration range. A remarkable response of 7520% was observed at 500 ppb, and even at 10 ppb, the sensor delivered a strong response of 46.5%, indicating exceptional sensitivity and low detection limit.

To further quantify the sensor's behavior, Figure 6(d) presents the calibration curve of the NbS$_2$/In$_2$O$_3$ heterostructure, plotting response versus NO$_2$ concentration. The curve was fitted with two linear segments: 10–100 ppb and 100–500 ppb. Both segments displayed excellent linearity, with adjusted R² values greater than 0.97, confirming that the sensor provides reliable and predictable output in both low and moderate NO$_2$ concentration ranges. Figure 6(e) shows the resistance transient curves of the heterostructure film during NO$_2$ exposure and recovery cycles. The sensor exhibited rapid and repeatable responses, with an average response time of ~22 seconds and a recovery time of ~155 seconds.

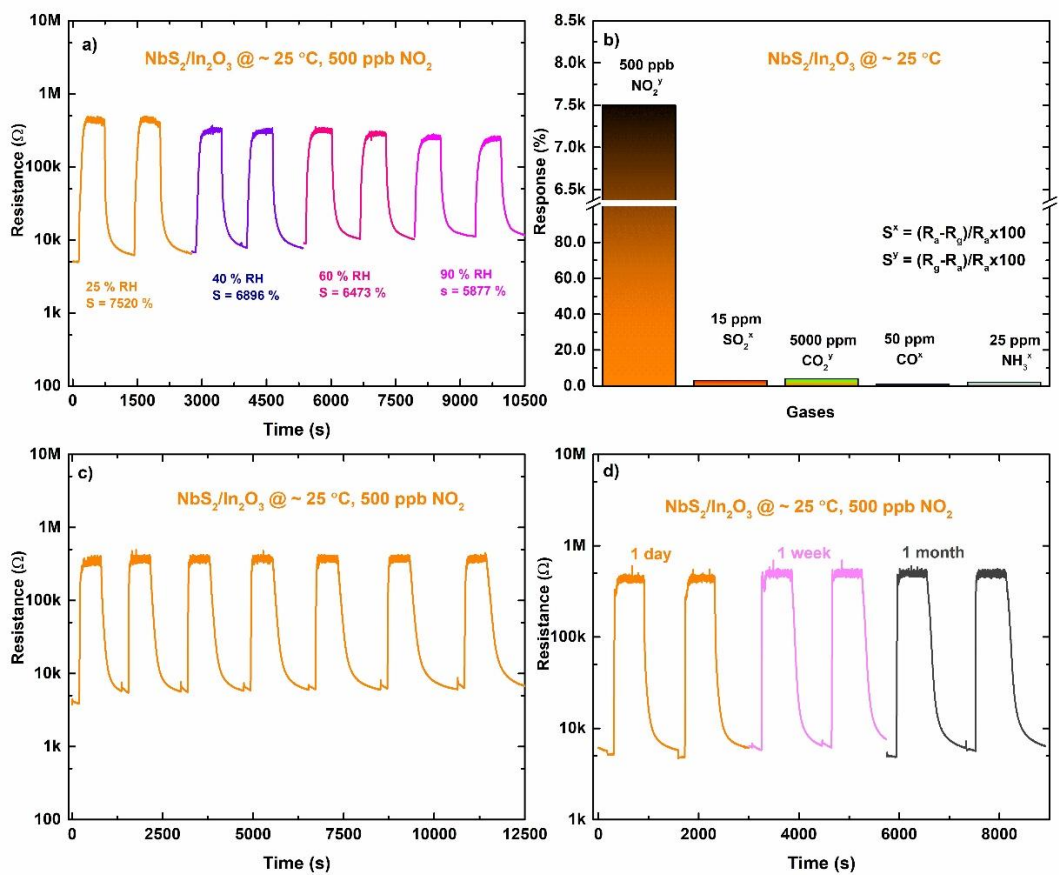

Figure 7. (a) NO$_2$ sensing performance of the heterostructure film under varying humidity levels, (b) selectivity against different interfering gases, (c) stability over continuous operation, and (d) repeatability of the sensing response

Figure 7(a) illustrates the sensing response of the heterostructure sensor under varying relative humidity (RH) conditions, ranging from 25% to 90% RH. As humidity increased, the sensor's baseline resistance progressively shifted to higher values. This trend suggests that water molecules adsorb onto the sensor surface, likely occupying or blocking active sites and thereby reducing charge carrier mobility. At 25% RH, the sensor exhibited a high response of approximately 7520%. However, the response gradually decreased with increasing humidity, reaching around 5877% at 90% RH. This behavior indicates competitive adsorption between NO$_2$ and H$_2$O molecules, where higher humidity levels inhibit the interaction of NO$_2$ with the sensing material. Despite the reduction in response with increased RH, the sensor maintained a relatively high response across all humidity levels, demonstrating its robustness and effectiveness under humid conditions.

Figure 7(b) demonstrates the selectivity of the NbS$_2$/In$_2$O$_3$ heterostructure sensor toward 500 ppb NO$_2$ when exposed to significantly higher concentrations of interfering gases: 15 ppm SO$_2$, 5000 ppm CO$_2$, 50 ppm CO, and 25 ppm NH$_3$. The stark contrast in response magnitudes underscores the sensor's exceptional specificity for NO$_2$, even in the presence of high levels of background interference. Further, figure 2c evaluates the repeatability of the NbS$_2$/ In$_2$O$_3$ sensor over multiple exposure cycles to 500 ppb NO$_2$ at ~25 °C, a critical parameter for assessing its reliability in practical applications. The resistance-time profile (figure 7(c)) confirms the sensor's ability to reproducibly detect NO$_2$ while maintaining stable and consistent performance across repeated gas pulses. Finally, figure 7(d) presents the sensor's long-term stability, showing that it retains a stable baseline resistance over a one-month period, with no noticeable drift at ~25 °C. These results demonstrate that the sensor maintains consistent sensing performance and baseline integrity over extended use.

**Gas sensing mechanism**

The proposed gas sensing mechanism involves charge transfer at the NbS$_2$/In$_2$O$_3$ heterojunction, where electrons migrate from semi metallic NbS$_2$ (with a lower work function) to semiconducting In$_2$O$_3$ (with a slightly higher work function), resulting in the formation of a depletion region at the heterointerface, as illustrated in Figure 8. In addition, gas adsorption at the surface of In$_2$O$_3$ leads to the formation of a surface depletion layer, which is a key feature in the sensing response. Ultraviolet photoelectron spectroscopy (UPS) data (Figure S5) provide insights into the electronic structure of the individual materials. As shown in Figure S5(a), the energy difference between the valence band maximum ($E_v$) and the Fermi level ($E_f$) for $NbS_2$ is approximately zero, confirming its semi metallic nature, which is further supported by its low resistance. In contrast, $In_2O_3$ exhibits an $E_v$–$E_f$ separation of approximately 2.62 eV, consistent with its n-type semiconducting behavior. The work functions of $NbS_2$ and $In_2O_3$, calculated from the secondary electron cutoff in UPS spectra (Figure S5(b)), are approximately 4.6 eV and 4.8 eV, respectively.

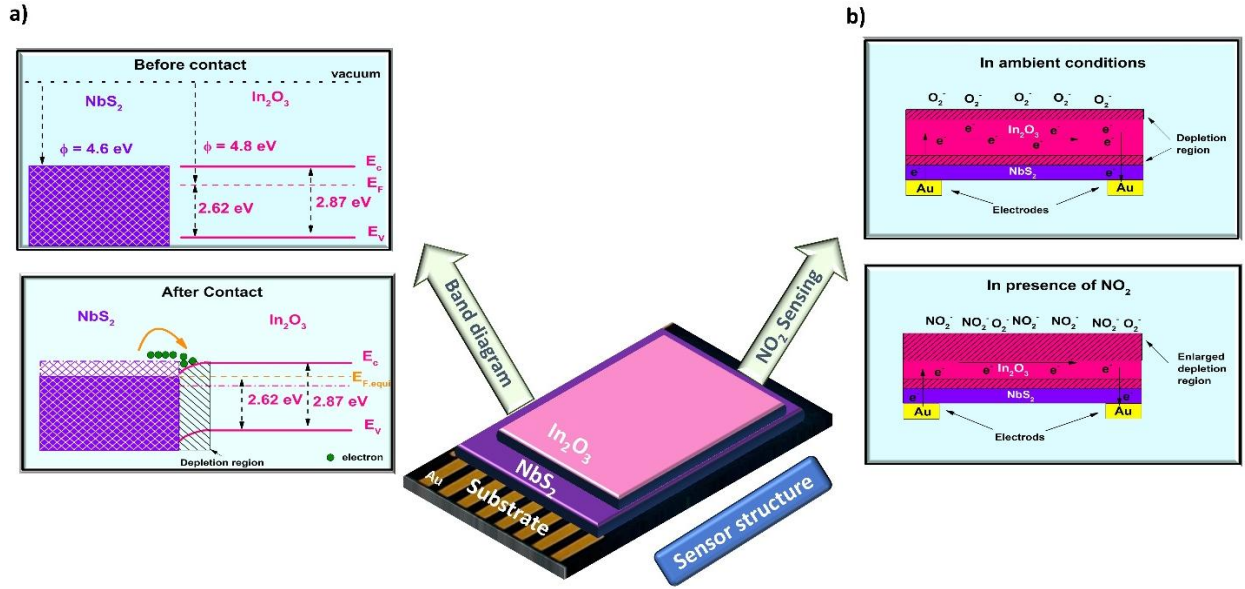

Figure 8 (a) Energy band diagram of the heterostructure film before and after contact, and (b) schematic of gas interactions with the heterostructure film in air and in the presence of $NO_2$.

Additionally, diffuse reflectance spectroscopy (DRS) combined with Tauc plot analysis (Figure S5(c)) yields an optical band gap of ~2.87 eV for $In_2O_3$, which aligns well with reported values. Figure 5(a) presents the schematic band alignment before contact. Upon heterojunction formation, electrons flow from $NbS_2$ to $In_2O_3$ to achieve Fermi level equilibration. This results in downward band bending in $In_2O_3$ near the interface, with the conduction band edge ($E_c$) shifting closer to the new equilibrium Fermi ($E_{f.equi}$) level due to increased electron concentration.

Furthermore, Figure 8(b) illustrates the sensor's behavior under ambient air and $NO_2$ exposure, showing both the direction of electron flow during resistance measurements and the formation of depletion regions. Under room temperature and ambient air conditions, oxygen molecules adsorb onto the surface of $In_2O_3$ and capture electrons from its conduction band to form chemisorbed oxygen species such as $O_2^-$ and $O^-$. This process induces a surface depletion layer, reducing the free carrier concentration in the near-surface region. In parallel, the heterojunction between $NbS_2$ and $In_2O_3$ results in additional electron transfer from $NbS_2$ to $In_2O_3$, creating a second depletion region at the interface. The combined effect of surface adsorption and interfacial charge redistribution results in a moderately low baseline resistance, around 8 k$\Omega$.

Upon exposure to $NO_2$, the gas molecules, being strong electron acceptors, interact with the $In_2O_3$ surface through reactions such as:

$$NO_{2(gas)} + e^- \leftrightarrow NO_{2(ads)}^- \qquad (2)$$

$$2NO_{2(gas)} + O_{2(ads)}^- + 2e^- \leftrightarrow 2NO_{2(ads)}^- + O_{(ads)}^- \qquad (3)$$

These reactions either replace pre-adsorbed oxygen species or occur in parallel with them, leading to additional electron withdrawal from the $In_2O_3$ conduction band. The depletion layer at the surface widens significantly due to the reduced free electron density, which in turn increases the potential barrier for charge carriers. This manifests as a substantial rise in sensor resistance, from the initial 8 k$\Omega$ to well above 100 k$\Omega$, often reaching several hundred kilo-ohms depending on $NO_2$ concentration and exposure duration.

The heterointerface with $NbS_2$ plays a crucial role in maintaining good electrical contact and facilitating rapid electron exchange due to its semi metallic nature, which enhances charge transfer kinetics. Moreover, the favorable work function alignment and band bending at the interface help in modulating carrier distribution and maximizing the sensor's dynamic response range. The synergistic combination of surface sensitivity (from $In_2O_3$), high charge mobility (from $NbS_2$), and the strong electron-withdrawing nature of $NO_2$ leads to excellent room-temperature sensing performance, characterized by high sensitivity and fast response.

**Conclusion**

In summary, we have demonstrated an $NbS_2/In_2O_3$ heterostructure-based $NO_2$ sensor with outstanding room-temperature performance. The sensor was fabricated by first synthesizing a $NbS_2$ film via CVD, transferring it onto a substrate, and subsequently spin-coating an $In_2O_3$ film on top. The formation of crystalline 2H-phase $NbS_2$ was confirmed by XRD and Raman spectroscopy. KPFM and XPS analyses further revealed charge transfer at the heterointerface from $NbS_2$ to $In_2O_3$. The resulting heterointerface exhibited dramatically enhanced gas sensing behavior compared to either material alone, achieving high sensitivity down to 10 ppb $NO_2$, reasonable response and recovery times, and excellent selectivity and long-term operational stability. The enhanced performance is attributed to the synergistic effect of surface adsorption on $In_2O_3$ and efficient charge transfer at the $NbS_2/In_2O_3$ interface, which forms a depletion region that significantly modulates carrier transport during $NO_2$ exposure. These findings underscore the utility of integrating metallic transition metal dichalcogenides with semiconducting metal oxides to create high-performance, room-temperature gas sensors suitable for practical environmental monitoring applications.

**CRediT authorship contribution statement**

**P.K. Shihabudeen:** Conceptualization, Methodology, Investigation, Data curation, Writing – original draft preparation, Writing– review & editing. **Alex Sam:** Conceptualization, Methodology, Investigation, Writing– review & editing. **Shih-Wen Chiu:** Methodology, Writing– review &

editing. **Ta-Jen Yen:** Resources**,** Writing– review & editing. **Kea-Tiong Tang:** Supervision, Writing –review & editing, Funding acquisition

**Supporting Information**

Supporting Information is available from the Wiley Online Library or from the author.

**Acknowledgment**

The Authors acknowledge the National Science and Technology Council (NSTC), Taiwan for financial support of the work (NSTC 113-2218-E-007 -019, and NSTC 113-2640-E-007 -005). We also acknowledge National Institutes of applied research- National center for instrumentation research for the CVD facility (NSTC 113-2119-M-492-001-MBK)

**Data Availability Statement**

The data that support the findings of this study are available from the corresponding author upon reasonable request.